\documentstyle[prl,aps,epsf]{revtex} 
\input epsf
\newcommand{\bb}{\begin{equation}}
\newcommand{\en}{\end{equation}}               

\draft
\begin{document}
\title{Charge--Fluctuation--Induced Non--analytic Bending Rigidity}
\author{A. W. C. Lau$\dag$ and P. Pincus$\dag\,\ddag$\\}
\address{$\dag$ Department of Physics and $\ddag$ Department of Materials, 
University of California Santa Barbara, CA 93106--9530}
\date{\today}
\maketitle
\begin{abstract}
In this Letter, we consider a neutral system of mobile positive and 
negative charges confined on the surface of curved films.  This may be an 
appropriate model for: i) a highly charged membrane whose counterions are 
confined to a sheath near its surface; ii) a membrane composed of 
an equimolar mixture of anionic and cationic surfactants in 
aqueous solution.  We find that the charge fluctuations contribute 
a non--analytic term to the bending rigidity that varies 
logarithmically with the radius of curvature.  This may lead to 
spontaneous vesicle formation, which is indeed observed in similar 
systems.
\end{abstract}

Electrostatics of charged objects such as polyelectrolytes and membranes
in aqueous solution plays an important role in many biological systems
\cite{inter}.    
The fundamental description of these systems has been the mean--field
approaches -- the Poisson--Boltzmann (PB) or Debye--H\"{u}ckel (DH) theory 
(for a review, see \cite{review}).  
However, for a highly charged surface, the Manning theory of counterion 
condensation \cite{cond}
provides an analytically tractable approximation to the 
PB theory.  Indeed, it has been demonstrated rigorously from the solutions 
to the PB equation \cite{soln}
that the electrostatic potential far away from
the charged surface is independent of the charge density above
a certain critical value, implying that
the counterions are confined to a thin layer close to the charged surface.  
However, like the PB theory, it fails to capture the correlation effects 
of the counterions since it expressly assumes that 
the ``condensed'' counterions are uniformly distributed.  
On physical grounds, we should expect that 
at low enough temperatures the fluctuations of these 
condensed counterions about a uniform density would give rise 
to new phenomena.  Indeed, recent 
simulations \cite{numerics,plane} show that the effective force between 
two like--charged rods and planar surfaces actually becomes attractive 
at short distances.  These suprising results shed new light on the 
understanding of the electrostatic adhesion between cells \cite{cell}
and the puzzling problem of DNA condensation \cite{rod}.  
In this Letter, we examine the effect of fluctuations of these condensed 
counterions on the bending rigidity of a charged membrane.  

The elastic properties of a fluid membrane are characterized 
by three macroscopic parameters -- a bending elastic modulus $\kappa$, 
a Gaussian modulus $\kappa_G$, and a spontaneous curvature $H_0$.
The deformation free energy per unit area, expressed in terms of 
the mean curvature $H$ and Gaussian curvature $K$ may be given by 
the Helfrich free energy \cite{hell,PB}: \bb
f = \frac{\kappa}{2}\, (H\, -\, H_0)^2 + \kappa_G\,K.
\label{hel}
\en
Within an additive constant, the free energy of a sphere with radius $R$ 
is given by $f_s = (2\,\kappa +  \kappa_G)/ R^2 - 2\kappa H_0/R$ and 
of a cylinder with radius $R$ by $ f_c =\kappa/2R^2 - \kappa H_0/R.$
Therefore, the parameters $\kappa H_0$ and $\kappa + \kappa_G$
may be determined from $f_s$ and $f_c$.

The problem of the electrostatic contribution to 
the bending constants of layered membranes within the PB mean field 
approach has been studied \cite{rigid}.  The electrostatic renormalization 
of the bending rigidity turns out to be positive; hence electrostatics 
augments the rigidity of charged membranes.  Here we go beyond these PB 
approaches by assuming that the surface charge density $n_0$ is 
sufficiently high 
that the condensed counterions are confined to a layer of thickness 
$\lambda << L$, where $\lambda$ is the Gouy--Chapman length, which scales
inversely with $n_0$ and $L$ is the linear size of the charged 
membrane.  By considering the in--plane fluctuations of the condensed 
counterions 
and charges on the membranes, we model the system effectively as
a 2--D coulomb gas interacting with a $r^{-1}$ potential.  This model
has yet another experimental realization -- a neutral membrane composed 
of a dilute mixture of anionic (--) and cationic (+) surfactants. 

The electrostatic free energy of the system is the sum of the entropy
of the charges and the electrostatic interaction energy among them:$$
\beta F_e = \sum_{i=\pm} \int d^{2}{\bf x} \,\,n_i({\bf x}) \{ \ln 
[n_i({\bf x})\,\lambda_T^{-2}] - 1 \}
+ \frac{l_B}{2} \sum_{i=\pm} \int\, d^{2}{\bf x} \! \int\,d^{2}{\bf x}'\,
\frac{n_i({\bf x})n_i({\bf x}')}
{|{\bf x}\,-\,{\bf x}'|} $$
\bb \;\;\;\;-\,l_B \int d^{2}{\bf x} \int d^{2}{\bf x}' \frac{n_{+}({\bf x})
n_{-}({\bf x}')}{|{\bf x}\,-\,{\bf x}'|}, 
\label{free}
\en
where $\lambda_T$ is the de Broglie wavelength of the charges,
$l_{B} \equiv \frac{e^{2}}{\epsilon k_{B}T} \approx 7 \AA$ is 
the Bjerrum length for an aqueous solution of dielectric constant 
$\epsilon = 80$ ($H_{2}O$), $\beta^{-1} \equiv k_B T$, $k_B$ is the 
Boltzmann constant, $T$ is the temperature, and $n_i({\bf x})$ is the 
coarse--grained density of the charges of species $i$.  The domain of 
the integral in Eq. (\ref{free}) spans the entire charged membrane.
In order to calculate the change in the free energy due to fluctuations, 
we assume that $n_i({\bf x}) = n_0 + \delta n_i({\bf x})$ 
and expand the electrostatic 
free energy to second order in $\delta n_i$\cite{pin}:\bb
\beta \Delta F_e= \frac{1}{2}\int d^{2}{\bf x}\,d^{2}{\bf x}
'\left[\frac{l_{B}}{|{\bf x}\,-\,{\bf x}'|} 
+ \frac{\delta^{2}({\bf x}\,-\,{\bf x}')}{2\,n_{0}} \right ]
\,\delta\sigma({\bf x})\,\delta\sigma({\bf x}'),
\label{variation}
\en 
where $\delta\sigma = \delta n_{+}\, -\, \delta n_{-}.$ 
The first term in the bracket is the Coulomb interaction of the charges. 
The second term comes from the second variation of 
the ideal gas entropy of the charges.  The change in the free energy 
is obtained by summing all fluctuations weighted by
the Boltzmann factor:\bb
\beta G_e = - \ln \left [ \int d\delta\sigma({\bf x})\,\exp{-\beta \Delta F } 
 \right ].
\label{energy}
\en
It should be mentioned that Eq. (\ref{energy}) contains a 
divergent self--energy term which has to be substracted out.
This means that we have to discard the first two terms in the 
expansion for $l_{B} \rightarrow 0$, as can be seen easily by
considering the zero temperature limit.  As $T \rightarrow 0$,
the free energy is reduced
to the electrostatic energy which is first order in $l_B$.  Since the 
self--energy is just a constant independent of temperature, it must be 
linear in $l_B$.  In the following, we employ this 
``substraction scheme'' together with Eq. (\ref{energy}) to calculate 
the free energy where charges are confined to the surfaces of 
three geometries: i) a plane, ii) a sphere, and iii) a cylinder.

For the case of charges confined to a plane $\Delta F$ in 
Eq. (\ref{variation}) can be diagonalized by Fourier transform and 
is quadratic in $\delta \sigma$.  Performing the Gaussian integrals 
in Eq. (\ref{energy}) and substracting out the self--energy term,
we obtain the free energy per unit area due to fluctuations 
\cite{pin,attard}\bb
\beta g_{pl} = 1/2 \int\,\frac{d^{2} {\bf q}}{(2 \pi)^{2}} \left \{
\ln \left [ 1 + \frac{1}{2\,|{\bf q}|\,\lambda_{D}} \right ] - 
\frac{1}{2\,|{\bf q}|\,\lambda_{D}}
\right \},
\label{plane}
\en
where $\beta = 1/k_B T$ and $1/\lambda_{D} = 8 \pi n_{0} l_{B}$, 
which scales like the Gouy--Chapman length, is a length scale analogous to
the Debye screening length in 3--D.
This result can also be obtained by solving the Debye--H\"{u}ckel equation 
in 2D \cite{raval}.  Note that Eq. (\ref{plane}) is ultravioletly divergent 
because of the infinite energy associated with the collapse of opposite 
charges. Thus a microscopic cut--off is necessary.  In ref.\cite{mayer}, 
the author shows by partial summation of the Mayer series that
the resulting free energy is convergent and indeed equivalent to a 
microscopic cut--off.  From Eq. (\ref{plane}), one can deduce 
that the screening of the charges is weak and the potential to 
the lowest order has the distance dependence of a dipolar field, where 
$\lambda_{D}$ plays the role of the dipole moment.

For the case of charges confined on a sphere of radius $R$, after following 
a similar procedure to that described above we obtain\bb
\beta g_{sp} = \frac{1}{8 \pi R^{2}} \sum_{l=0}^{\infty} (\,2l+1\,)
\left\{ \ln \left [ 1 + \frac{R/\lambda_{D}}{2l + 1} \right ] 
-  \frac{R/\lambda_{D}}{2l + 1} \right \}.
\label{sphere}
\en 
It is easy to show that by setting $k=l/R$ and taking the limit 
$ R \rightarrow \infty$, we recover the planar result.  Equivalently
we may write Eq. (\ref{plane}) as\bb
\beta g_{pl} = \frac{1}{8 \pi R^{2}} \int_{-1/2}^{\infty} dl\, (\,2l+1\,)
\left\{ \ln \left [ 1 + \frac{R/\lambda_{D}}{2l + 1} \right ] 
-  \frac{R/\lambda_{D}}{2l + 1} \right \}.
\label{plane2}
\en
The difference $g_{sp} - \,g_{pl},$ can be evaluated 
as an asymptotic expansion in $1/R$ using the Euler--MacLaurin 
summation formula \cite{wat} with $f(l) = 
(2l + 1) \ln ( 2l + 1 + R/\lambda_{D})$.  The result is:\bb
\beta\,(\,g_{sp} - \,g_{pl}\,) = - \frac{11}{96 \pi R^{2}}\,\ln
(\,R/\lambda_{D}\,) + ....
\label{result}
\en
In deriving the result above, we have regularized the 
integral in Eq. (\ref{plane2}) and the sum in Eq. (\ref{sphere}) by 
an ultraviolet cut--off $\Lambda$.  However, the leading term in 
Eq. (\ref{result}) is cut--off independent 
and those higher order cut--off dependent terms tend to zero as $\Lambda 
\rightarrow \infty$.

For the case of a cylinder, we obtain the free energy:\bb
\beta g_{cyl} = \frac{1}{4 \pi R} \sum_{ m \geq 0} \int_{0}^{\infty}\,dq\, 
\frac{2}{\pi}\, \left \{ \ln \left [  1 + 
\frac{R}{\lambda_{D}}\,I_m(qR)K_m(qR) \right ]  
- \frac{R}{\lambda_{D}}\,I_m(qR)K_m(qR) \right \},
\label{cylinder}
\en
where $I_m$ and $K_m$ are modified Bessel functions of order $m$. 
The evaluation of the integrals here is relatively difficult.
However, we argue that $g_{cyl} - \,g_{pl}$ has the following asymptotic 
expansion:\bb
\beta\,(\,g_{cyl} - \,g_{pl}\,) =
- \frac{1}{48 \pi R^{2}}\,\ln (\,R/\lambda_{D}\,) + ...,
\label{cyl}
\en
for $R \rightarrow \infty$.  First, we note that the only relevent 
contributions to the $q$--integral in 
Eq. (\ref{cylinder}) are sharply peaked at $q \approx 0$ with width 
$\Delta\,q \approx m/R$.  Hence, the Bessel functions can be approximated by
$ I_m(qR)K_m(qR) \sim 1/2m$, yielding \bb
\beta g_{cyl} = \frac{1}{4 \pi R^2} \sum_{ m \geq 0}\,m
\left [ \ln \left (  1 + \frac{R/\lambda_{D}}{2m} \right ) 
- \frac{R/\lambda_{D}}{2m} \right] + O(1/R^3).
\en
Equation (\ref{cyl}) can now be obtained by using the 
Euler--MacLaurin summation formula with $f(m) = m \log( 2m + R/\lambda_{D}).$

The modificaitons to the bending constants can be obtained from 
Eqs. (\ref{result}) and (\ref{cyl}) to yield
\begin{eqnarray}
&&\Delta\kappa = - \frac{1}{24 \pi}\, \ln ( R/\lambda_{D}),\nonumber\\
& &\Delta\kappa_G = - \frac{1}{ 12\,\pi}\,\ln ( R/\lambda_{D}).
\label{renormalization}
\end{eqnarray}
We thus find that the contribution to the membrane elastic contants due to 
charge fluctuations is non--analytic.  This kind of 
non--analyticity in the bending constants exists in the literature
in other situations, for example in a system consisting of a membrane 
and rod--like cosurfactants\cite{non}.  In the present case, this 
non--analyticity can be considered a signiture of 2--D charged systems.  
The DH theory in 3--D yields an expression for the change in the free 
energy per unit volume \cite{landau}
$ \Delta f_{3D} \sim -\lambda_{D}^{-3} + ....$  In constrast, 
Eq. (\ref{plane}) has a similar expansion for the free energy per unit 
area but contains a logarithmic term\cite{raval}: 
$\,\,\Delta f_{2D} \sim - \lambda_{D}^{-2} \ln[\lambda_{D}/a] + ....$  
Therefore, it is not unexpected to find logarithmic corrections 
to the bending constants.  

Secondly, we remark that both $\Delta\kappa$ and $\Delta\kappa_G$
are negative, in contrast to the mean--field PB contributions, where
the renormalization of the bending moduli are always positive and 
the Gaussian moduli may be negative in some cases.  
In a system in which $R/\lambda_{D} >> 1$,  $\Delta\kappa$ is large 
compared to the mean--field contribution and the membrane becomes more 
flexible.  Therefore, charge fluctuations induce bending of a charged 
membrane.  This conclusion can be seen physically in the light of the recent 
works \cite{pin,attard} on attractive interactions between two planar
charge--fluctuating membranes.  It is found that for
large distance $h$ separating the two membranes, the attractive force 
per unit area scales as $h^{-3}$.  Now, a sphere or a cylinder can be 
approximated as two flat planar surfaces in the limit $R \rightarrow \infty$
and their interaction free energies per unit area therefore 
should scale like $f \sim -\,R^{-2}$.  Hence negative contribution 
to the bending modulus.

The negative contribution of $\Delta\kappa_G$ from charge 
fluctuations can also be understood physically as follows.  
Since the $r^{-1}$ potential is rotationally invariant, one
might expect that the charged surface has rotational symmetry
when the the electrostatic free energy is a minimum.
Now, recall that the Gaussian curvature, expressed 
in terms of the principal radii $R_1$ and $R_2$,
is $K = \frac{1}{R_1\,R_2}$.  Therefore, as can be easily seen,
the free energy in Eq. (\ref{hel}) is indeed lowest 
(since $\Delta\kappa_G <\,0$) if there is rotational 
symmetry about an axis normal to the charged surface or $R_1 = R_2$, 
in accordance with the rotational symmetry of the $r^{-1}$ potential.  
Furthermore, the fact that $\Delta\kappa_G < 0$ 
has interesting experimental consequences since strongly 
negative values of $\kappa_G$ favor the formation of many disconnected pieces 
with no rims, like spherical vesicles.  Therefore, when the 
surface charge density is made sufficiently large, the membrane 
might spontaneously form vesicles, due to fluctuations
of condensed counterions.
Experiments \cite{vesi} on charged surfactant systems supports
this conclusion.

The result presented in this Letter is particularly relevant to 
recent experiments \cite{vesicle} where the authors find the formation of 
vesicles by mixing anionic and cationic surfactants.  
Two aspects of their experiment can be qualitatively accounted for by the 
present model.  They find, in equilibrium, large vesicles with $R \sim 
1000\,\,\AA$ and substantial size polydispersity.  Indeed, the vesicle 
free energy per unit area given by\bb
f_{ves} = \kappa_b/R^2 - \frac{11}{96 \pi R^{2}}\,\ln
(\,R/\lambda_{D}\,),
\label{ves}
\en
where $\kappa_b$ is the bare value of the bending rigidity, 
has an equlibrium value $R^* \sim \lambda_{D}\exp(\kappa_b/k_B T)$,
which can be large even for a moderate value of $\kappa_b$ of 
the order of 3--5 $k_B T$.  Furthermore, the second derivative of the free 
energy $f''(R^*) \sim e^{-\kappa_b/k_B T}$ is exponentially small. 
Hence the variance or fluctuations in $R$, 
$< (\Delta R)^2 > \sim 1/f''(R^*)$  is large, implying 
size polydispersity.

In conclusion, by studying fluctuations of charges on curved
films, we have deduced non--analytic contributions to the
bending energy of a membrane.  Our calculation is applicable
to condensed counterions on a highly charged membrane, and 
mixing of surfactants of opposite charges.  For the latter case,
we find qualitative agreements with experiments.  

We would like to thank S. Safran, G. Rossi, and C. Safinya 
for stimulating and helpful discussions.  AL and PP acknowledge 
support from NSF grants MRL--DMR--9632716, DMR--9624091, and DMR--9708646.
\pagebreak

\end{document}